# Nonreciprocal and chiral single-photon scattering for giant atoms

Yao-Tong Chen[1], Lei Du [1,2], Lingzhen Guo [3,4], Zhihai Wang[1], Yan Zhang [1,5✉], Yong Li [2,6✉] & Jin-Hui Wu [1✉]

Quantum optics with giant atoms has provided a new paradigm to study photon scatterings. In this work, we investigate the nontrivial single-photon scattering properties of giant atoms being an effective platform to realize nonreciprocal and chiral quantum optics. For two-level giant atoms, we identify the condition for nonreciprocal transmission: the external atomic dissipation is further required other than the breaking of time-reversal symmetry by local coupling phases. Especially, in the non-Markovian regime, unconventional revival peaks periodically appear in the reflection spectrum. To explore more interesting scattering behaviors, we extend the two-level giant-atom system to Δ-type and ∇-type three-level giant atoms coupled to double waveguides with different physical mechanisms to realize nonreciprocal and chiral scatterings. Our proposed giant-atom structures have potential applications of high-efficiency targeted routers that can transport single photons to any desired port deterministically and circulators that can transport single photons between four ports in a cyclic way.

[1] School of Physics and Center for Quantum Sciences, Northeast Normal University, Changchun 130024, China. [2] Center for Theoretical Physics and School of Science, Hainan University, Haikou 570228, China. [3] Department of Applied Physics, Nanjing University of Science and Technology, Nanjing 210094, China. [4] Max Planck Institute for the Science of Light, 91058 Erlangen, Germany. [5] National Demonstration Center for Experimental Physics Education, Northeast Normal University, Changchun 130024, China. [6] Synergetic Innovation Center for Quantum Effects and Applications, Hunan Normal University, Changsha 410081, China. ✉email: zhangy345@nenu.edu.cn; yongli@hainanu.edu.cn; jhwu@nenu.edu.cn

Waveguide quantum electrodynamics (QED) studies the interactions between atoms and one-dimensional waveguide modes, providing an excellent platform for constructing long-range interactions and engineering large-scale quantum networks[1–5]. In experiments, typical candidates of implementing waveguide QED systems include quantum dots coupled to photonic crystal waveguides[6,7], superconducting qubits coupled to transmission lines[8–11], ultracold atoms coupled to optical fibers[12,13], etc. To date, waveguide QED has inspired a number of exotic phenomena, such as atom-like mirrors[14,15], dynamical Casimir effects[16], single-photon routing[17–19], and bound states in the continuum[20].

In general, the atom can be viewed as a point when coupled with the waveguide due to its negligible size compared to the wavelength of waveguide modes. Nevertheless, in a recent experiment, a superconducting transmon qubit was designed to interact with surface acoustic waves via multiple coupling points whose separation distances can be much larger than the wavelength of the waves[21]. Instead, a generalized theory called "giant atom" has been developed to describe such situations[22]. Since the first theoretical study in 2014[23], the giant-atom scheme has been broadly investigated with superconducting qubits[24–28], coupled waveguide arrays[29], and cold atoms[30]. With such nonlocal coupling schemes, a series of tempting quantum phenomena have been demonstrated, including frequency-dependent relaxation rate and Lamb shift[23,27,31], non-exponential atomic decay[24,25], decoherence-free interatomic interaction[27,32,33], exotic bound states[26,34], modified topological effects[35], and quantum Zeno and quantum anti-Zeno effects[36]. Giant atoms have emerged as a new paradigm in quantum optics and require more comprehensive understanding in physics.

On the other hand, controlling the flow of photons, especially realizing asymmetric photonic propagations in waveguide QED systems, is crucial for constructing nonreciprocal optical devices[37–42]. To this end, one could break the time-reversal symmetry of the system such that the interactions between the atoms and the waveguide modes are direction-dependent[18,43,43–47]. Such a paradigm, also known as chiral quantum optics[43], can be achieved via several methods, such as the spin-momentum locking effect[48–50], inserting circulators in superconducting circuits[51–53], applying topological waveguides[54,55], synthesizing artificial gauge fields[56], adding spin-orbit coupling to Bose-Einstein quasicondensates[57], and using Rydberg atoms or trapped ions[58]. Based on the chiral interaction, targeted photonic routers[19], single-photon circulators[59,60], cascaded quantum networks[61–63], and enhanced entanglement[64,65] have been proposed. Recently, the concept of giant atom has been introduced to chiral quantum optics, making some advanced functionalities possible, such as chiral bound states[34], dark states without coherent drives[33], and non-Markovianity induced nonreciprocity[66]. These seminal works inspire us to explore more intriguing effects in chiral giant-atom setups, especially with multi-level structure[66–68].

In this paper, we investigate how external atomic dissipations outside the waveguide and local coupling phases affect the single-photon scattering properties of a two-level giant atom with two atom-waveguide coupling points. By taking into account the phase difference between two coupling points, we find that the giant atom behaves like a chiral small atom in the Markovian regime but exhibits peculiar giant-atom effects in the non-Markovian regime. We physically demonstrate that the breaking of time-reversal symmetry by local coupling phases is not sufficient for realizing nonreciprocal photon scatterings. In fact, in the absence of the external atomic dissipation, the scatterings are always reciprocal even if the atomic spontaneous emission becomes chiral[68,69]. In order to realize asymmetric scattering for a giant atom without external dissipation, we propose a ∇-type giant atom coupled to two waveguides. In such way, we realize the nonreciprocal and chiral scatterings with single ∇-type atom. Targeted routing and circulation schemes can also be realized via such scatterings with proper phases. Finally, we consider a Δ-type giant atom and compare its properties with that of ∇-type one. We reveal that, the nonreciprocal scatterings stem from the quantum interference effect in the closed-loop atom-level structure for the Δ-type giant atom, but from the nontrivial coupling phase difference for the ∇-type giant atom.

## Results and discussion

**Two-level giant atom coupled to a single waveguide.** As schematically shown in Fig. 1a, we consider a two-level giant atom coupled to a waveguide at two separated points $x=0$ and $x=d$. The atom-waveguide coupling coefficients are $ge^{i\theta_1}$ and $ge^{i\theta_2}$, respectively, with local coupling phases $\theta_1$ and $\theta_2$ for inducing some intriguing interference effects to the scattering properties as will be discussed below. With superconducting quantum devices, the local coupling phases can be introduced with Josephson-junction loops threaded by external fluxes[69].

Under the rotating wave approximation, the real-space Hamiltonian of the model can be written as ($\hbar=1$ hereafter)

$$
\begin{aligned}
H &= H_w + H_a + H_I, \\
H_w &= \int_{-\infty}^{+\infty} dx \left[ a_L^\dagger(x)\left(\omega_0 + iv_g \frac{\partial}{\partial x}\right) a_L(x) \right.\\
&\quad \left. + a_R^\dagger(x)\left(\omega_0 - iv_g \frac{\partial}{\partial x}\right) a_R(x) \right], \\
H_a &= \left(\omega_e - i\frac{\gamma_e}{2}\right)|e\rangle\langle e|, \\
H_I &= \int_{-\infty}^{+\infty} dx \Big\{ \delta(x) g e^{i\theta_1} \left[ a_R^\dagger(x) + a_L^\dagger(x) \right] |g\rangle\langle e| \\
&\quad + \delta(x-d) g e^{i\theta_2} \left[ a_R^\dagger(x) + a_L^\dagger(x) \right] |g\rangle\langle e| + \text{H.c.} \Big\}.
\end{aligned}
\tag{1}
$$

Here $H_w$ represents the free Hamiltonian of the waveguide modes with $v_g$ being the group velocity of photons in the waveguide. $a_{R,L}$ ($a_{R,L}^\dagger$) are the bosonic annihilation (creation) operators of the right-going and left-going photons in the waveguide, respectively;

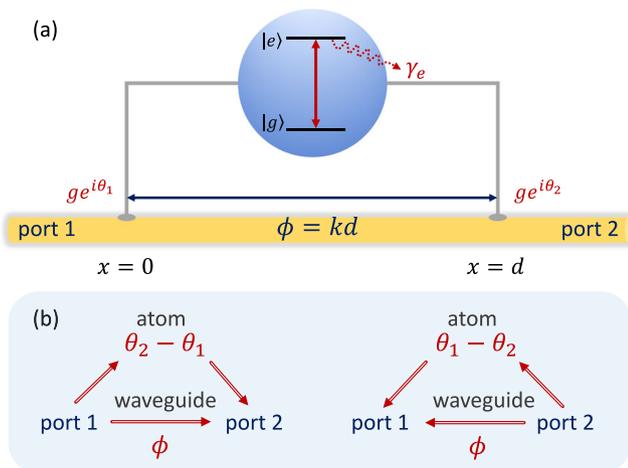

**Fig. 1 Schematic representation of the model and the photon paths. a** A two-level giant atom coupled to a waveguide at $x=0$ and $x=d$, respectively, with individual local coupling phases $\theta_{1,2}$. **b** Two paths of a single photon propagating from port 1 to port 2 (left) or from port 2 to port 1 (right).

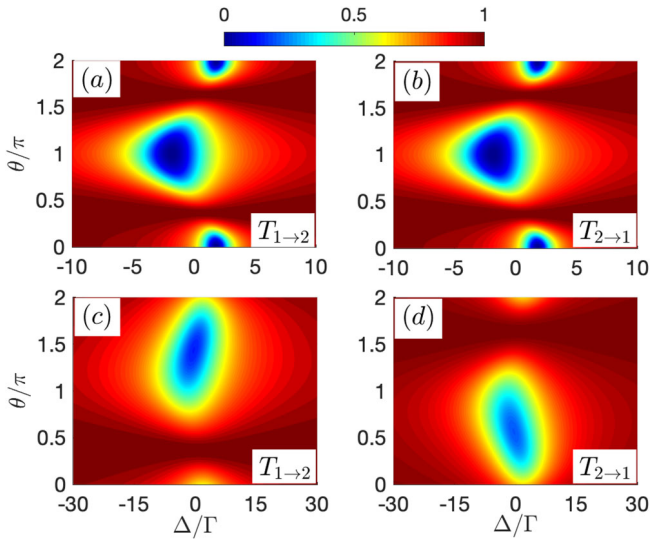

**Fig. 2 Reciprocal and nonreciprocal transmission behaviors in a two-level giant-atom system.** Transmission rates $T_{1\to 2}$ and $T_{2\to 1}$ versus the detuning $\Delta$ and the phase difference $\theta$. **a** and **b** with the atomic dissipation rate $\gamma_e = 0$; **c** and **d** with $\gamma_e/\Gamma = 10$, where $\Gamma$ is the atomic emission rate into the waveguide. Other parameters: accumulated phase $\phi_0 = \pi/2$ and photon propagation time $\tau\Gamma = 0.01$.

$\omega_0$ is the frequency around which the dispersion relation of the waveguide mode is linearized[1,70]. $H_a$ is for the atom, where $\omega_e$ describes the transition frequency between the ground state $|g\rangle$ and the excited state $|e\rangle$; $\gamma_e$ is the external atomic dissipation rate due to the non-waveguide modes in the environment. $H_I$ describes the interactions between the atom and the waveguide, where the Dirac delta functions $\delta(x)$ and $\delta(x-d)$ indicate that the atom-waveguide couplings occur at $x=0$ and $x=d$, respectively.

Considering that the total excitation number is conserved in rotating wave approximation, the eigenstate of the system can be expressed in the single-excitation subspace as

$$|\Psi\rangle = \int_{-\infty}^{+\infty} dx [\Phi_R(x) a_R^\dagger(x) + \Phi_L(x) a_L^\dagger(x)] |0,g\rangle + u_e |0,e\rangle, \quad (2)$$

where $\Phi_{R,L}(x)$ are the density of probability amplitudes of creating the right-going and left-going photons at position $x$, respectively; $u_e$ is the excitation amplitude of the atom; $|0,g\rangle$ denotes the vacuum state of the system. The probability amplitudes can be determined by solving the eigenequation $H|\Psi\rangle = E|\Psi\rangle$, which leads to

$$E\Phi_R(x) = \left(\omega_0 - iv_g \frac{\partial}{\partial x}\right) \Phi_R(x) + g\left[e^{i\theta_1}\delta(x) + e^{i\theta_2}\delta(x-d)\right] u_e,$$

$$E\Phi_L(x) = \left(\omega_0 + iv_g \frac{\partial}{\partial x}\right) \Phi_L(x) + g\left[e^{i\theta_1}\delta(x) + e^{i\theta_2}\delta(x-d)\right] u_e, \quad (3)$$

$$Eu_e = \left(\omega_e - i\frac{\gamma_e}{2}\right) u_e + ge^{-i\theta_1}[\Phi_R(0) + \Phi_L(0)] + ge^{-i\theta_2}[\Phi_R(d) + \Phi_L(d)].$$

Assuming that a photon with the renormalized wave vector $k$ satisfying the linear dispersion relation $E = \omega_0 + kv_g$ is incident from port 1 of the waveguide, the wave functions $\Phi_{R,L}(x)$ can be written as

$$\Phi_R(x) = e^{ikx}\{\Theta(-x) + A[\Theta(x) - \Theta(x-d)] + t\Theta(x-d)\},$$

$$\Phi_L(x) = e^{-ikx}\{r\Theta(-x) + B[\Theta(x) - \Theta(x-d)]\}, \quad (4)$$

where $\Theta(x)$ is the Heaviside step function. Here, $t$ and $r$ denote the single-photon transmission and reflection amplitudes in the regions of $x > d$ and $x < 0$, respectively. We define $A$ and $B$ as the probability amplitudes for the right-going and left-going photons between the two coupling points ($0 < x < d$), respectively.

Substituting Eq. (4) into Eq. (3), we obtain

$$\begin{aligned}
0 &= -iv_g(A-1) + ge^{i\theta_1} u_e, \\
0 &= -iv_g(t-A)e^{i\phi} + ge^{i\theta_2} u_e, \\
0 &= -iv_g(r-B) + ge^{i\theta_1} u_e, \\
0 &= -iv_g Be^{-i\phi} + ge^{i\theta_2} u_e, \\
0 &= \frac{g}{2} e^{-i\theta_1}(A+B+r+1) + \frac{g}{2} e^{-i\theta_2} \\
  &\quad \times (Ae^{i\phi} + Be^{-i\phi} + te^{i\phi}) - \left(\Delta + i\frac{\gamma_e}{2}\right) u_e
\end{aligned} \quad (5)$$

with $\Delta = E - \omega_e$ being the detuning between the incident photon and the atomic transition. In the case of near resonant couplings with $E \sim \omega_e$ (i.e., $|\Delta/\omega_e| \ll 1$), the transmission and reflection amplitudes can be obtained from solving Eq. (5) as

$$t = \frac{\Delta + i\frac{\gamma_e}{2} - 2\Gamma e^{i\theta}\sin\phi}{\Delta + i\frac{\gamma_e}{2} + 2i\Gamma(1 + e^{i\phi}\cos\theta)}, \quad (6a)$$

$$r = \frac{[2i\Gamma(1 + e^{i\phi}\cos\theta) + 2\Gamma e^{i\theta}\sin\phi][1 + e^{i(\theta+\phi)}]}{[\Delta + i\frac{\gamma_e}{2} + 2i\Gamma(1 + e^{i\phi}\cos\theta)][1 + e^{i(\theta-\phi)}]}, \quad (6b)$$

where $\theta = \theta_2 - \theta_1$ is the phase difference between the two atom-waveguide coupling channels and $\Gamma = g^2/v_g$ is the rate of the atomic emission into the waveguide. Compared with the setup of a two-level small atom coupled locally to a waveguide, such giant atom shows phase-dependent effective detuning and decay rate given by $\Delta - 2\Gamma\cos\theta\sin\phi$ and $\gamma_e/2 + 2\Gamma(1 + \cos\theta\cos\phi)$, respectively[23]. In fact, a left-incident (right-incident) photon can propagate from $x=0$ to $x=d$ (from $x=d$ to $x=0$) via two different paths: it can either keep on propagating along the waveguide, or be absorbed at $x=0$ ($x=d$) and re-emitted at $x=d$ ($x=0$) by the atom, as shown in Fig. 1b. For the left-incident photon, the two paths yield phase accumulations $\phi$ and $\theta$, respectively, which determine the phase-dependent interference effect jointly.

For the right-incident photon, the propagation process is equivalent to that of the left-incident one yet with exchanged coupling phases, i.e., $\theta_1 \leftrightarrow \theta_2$. Therefore, the transmission and reflection amplitudes for the right-incident photon are expressed as

$$t' = \frac{\Delta + i\frac{\gamma_e}{2} - 2\Gamma e^{-i\theta}\sin\phi}{\Delta + i\frac{\gamma_e}{2} + 2i\Gamma(1 + e^{i\phi}\cos\theta)}, \quad (7a)$$

$$r' = \frac{[2i\Gamma(1 + e^{i\phi}\cos\theta) + 2\Gamma e^{-i\theta}\sin\phi][1 + e^{-i(\theta-\phi)}]}{[\Delta + i\frac{\gamma_e}{2} + 2i\Gamma(1 + e^{i\phi}\cos\theta)][1 + e^{-i(\theta+\phi)}]}, \quad (7b)$$

which are also consistent with the results obtained by rewriting the wave functions for the right-incident photon. In addition, it is worth noting that the accumulated phase of a propagating photon can be written as $\phi = (k_0 + k)d = \phi_0 + (E - \omega_0)\tau = \phi_0 + \tau\Delta$ with $\phi_0 = k_0 d$ and $\tau = d/v_g$ by taking $\omega_0 = \omega_e$ for convenience. As usual, we have discarded $k_0$ in $H_w$ of Eq. (1) and $\Phi_{L,R}$ of Eq. (4) without changing other equations, and will take $\phi_0 = \alpha$ to replace $\phi_0 = \alpha + 2m\pi$ with $m$ being a positive integer and $0 \le \alpha < 2\pi$ in the following discussions. Hence, it is viable to work in the Markovian regime with $|\tau\Delta| \sim \tau\Gamma \ll 1 \sim \phi_0$ when $d$ is not too large, while in the non-Markovian regime with $|\tau\Delta| \sim \tau\Gamma \sim 1 \sim \phi_0$ when $d$ is large enough. For a transmon qubit considered here, $\omega_e$ and $\Gamma$ are of the order of GHz and 0.1MHz[25,27,28], respectively, ensuring thus the validity of rotating wave approximation mentioned before Eqs. (1) and (2).

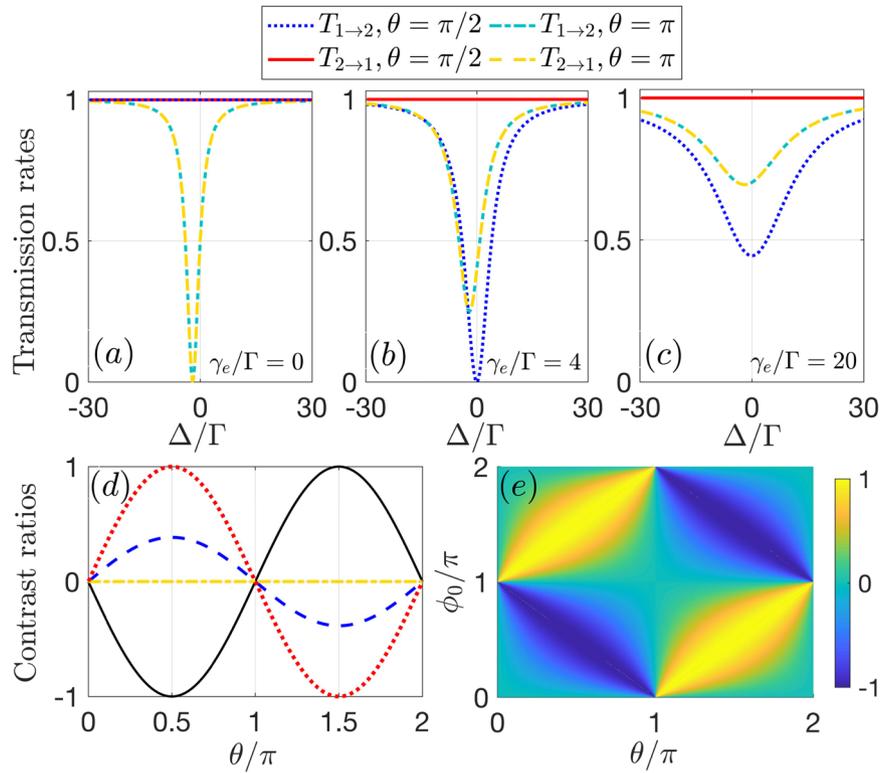

**Fig. 3 The influence of the coupling phase difference and accumulated phase on transmission rates and contrast ratios.** Transmission rates $T_{1\to 2}$ and $T_{2\to 1}$ versus the detuning $\Delta$ with accumulated phase $\phi_0 = \pi/2$ and different atomic dissipation rates **a** $\gamma_e/\Gamma = 0$; **b** $\gamma_e/\Gamma = 4$; **c** $\gamma_e/\Gamma = 20$, where $\Gamma$ is the atomic emission rate into the waveguide. **d** Contrast ratios $I$ and $D$ versus the coupling phase difference $\theta$ with $\phi_0 = \pi/2$ and $\Delta = 0$. The yellow dot-dashed, red dotted, and blue dashed lines are $I$ with $\gamma_e/\Gamma = 0$, $\gamma_e/\Gamma = 4$, and $\gamma_e/\Gamma = 20$, respectively, and the black solid one represents $D$ independent of $\gamma_e$. **e** Contrast ratio $D$ versus $\theta$ and $\phi_0$ with $\Delta = 0$. Other parameters: photon propagation time $\tau\Gamma = 0.01$.

*Reciprocal and nonreciprocal transmissions.* We first focus on the Markovian regime of $\tau \ll 1/(2\Gamma + \gamma_e/2)$, where $\phi \approx \phi_0$ according to the Taylor expansion because this substitution gives correct Lamb shift and modified emission rate in the Markovian limit[24,29]. In Fig. 2, we plot the transmission rates $T_{1\to 2} = |t|^2$ and $T_{2\to 1} = |t'|^2$ as functions of the detuning $\Delta$ and the phase difference $\theta$ with and without external atomic dissipations. Owing to the interference between two photon paths mentioned above, the scattering behavior changes periodically with $\theta$. For $\gamma_e = 0$ as shown in Fig. 2a, b, the single-photon scattering is reciprocal, i.e., $T_{1\to 2} \equiv T_{2\to 1}$, although the time-reversal symmetry is broken due to the nontrivial phase difference $\theta$ arising from the interference.

This counterintuitive phenomenon can be explained by comparing Eqs. (6a) and (7a). On one hand, the transmission amplitudes $t$ and $t'$ share the same denominator that is an even function of $\theta$. On the other hand, the numerators of $t$ and $t'$ in Eqs. (6a) and (7a) can be rewritten as

$$\begin{aligned}\Delta - 2\Gamma\sin\phi\cos\theta + i\left(\frac{\gamma_e}{2} - 2\Gamma\sin\phi\sin\theta\right),\\ \Delta - 2\Gamma\sin\phi\cos\theta + i\left(\frac{\gamma_e}{2} + 2\Gamma\sin\phi\sin\theta\right).\end{aligned} \quad (8)$$

Equation (8) clearly shows that nonreciprocal single-photon transmissions ($|t|^2 \ne |t'|^2$) can be achieved only if a finite external atomic dissipation rate is taken into account ($\gamma_e > 0$). This can be observed by the transmission spectra shown in Fig. 2c, d.

When $\gamma_e = 0$, Fig. 3a depicts the transmission rates $T_{1\to 2}$ and $T_{2\to 1}$ versus the detuning $\Delta$ with various $\theta$. For $\theta = \pi/2$, we find $T_{1\to 2} = T_{2\to 1} \equiv 1$ over the whole range of the detuning, implying that reflections are prevented for both directions. For $\theta = \pi$, however, the transmission spectrum exhibits the Lorentzian line shape with phase-dependent Lamb shift and linewidth (decay rate)[23]. In both cases ($\theta = \pi/2, \pi$), the transmissions are reciprocal, yet the atomic excitation probabilities are different as will be discussed below. When $\gamma_e \ne 0$, as shown in Fig. 3b and (c), the scattering becomes nonreciprocal if $\theta = \pi/2$; however, with $\theta = \pi$, the scatterings are still reciprocal even in the presence of the external dissipation. We also can see from the three curves corresponding to $\theta = \phi_0 = \pi/2$ a non-monotonic behavior of transmission rate $T_{1\to 2}$ (in particular, $T_{1\to 2} = 0$ at $\Delta = 0$ for $\gamma_e/\Gamma = 4$) with the increase of external decay rate $\gamma_e$ while $T_{2\to 1}$ remains unity independent of $\Delta$ and $\gamma_e$. This can be understood by resorting to Eqs. (6a) and (7a) restricted by $\theta = \phi_0 = \pi/2$, from which it is easy to find that $T_{2\to 1} = |t'|^2 \equiv 1$ while $T_{1\to 2} = |t|^2$ exhibits a vanishing (nonzero) minimum for $\Delta = 0$ ($\Delta \ne 0$) at the optimal $\gamma_e = 2\sqrt{\Delta^2 + 4\Gamma^2}$ as determined by setting $\partial T_{1\to 2}/\partial \gamma_e = 0$. Physically, this is an interference result of two propagating paths. In Eq. (6a), the direct path denoted by $-2\Gamma e^{i\theta}\sin\phi$ is along the waveguide from $x = 0$ to $x = d$; the indirect path denoted by $\Delta + i\gamma_e/2$ is via an absorption at $x = 0$ and an emission at $x = d$. The two paths will contribute a perfect destructive interference leading to $t = 0$ in the case of $\sin\phi = \pm\sqrt{\gamma_e^2 + 4\Delta^2}/4\Gamma$ and $\tan\theta = \gamma_e/2\Delta$ indicating that $T_{1\to2}$ can also exhibit a vanishing minimum for $\Delta \neq 0$ if we have $\sqrt{\gamma_e^2 + 4\Delta^2} \leq 4\Gamma$. A similar analysis on Eq. (7a) shows that a perfect destructive interference leading to $t' = 0$ will occur in the case of $\sin\phi = \pm\sqrt{\gamma_e^2 + 4\Delta^2}/4\Gamma$ and $\tan\theta = -\gamma_e/2\Delta$ due to a reversed phase difference $(\theta \to -\theta)$ in the direct path. Hence, it is impossible to simultaneously have $T_{1\to2} = 0$ and $T_{2\to1} = 0$ for a nonzero external decay $(\gamma_e \neq 0)$.

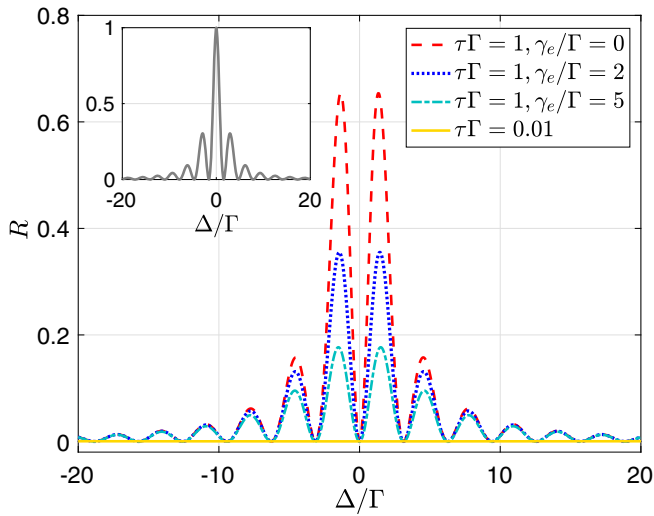

**Fig. 4 Reflection behaviors in the Markovian and non-Markovian regime.** Reflection rate $R$ versus the detuning $\Delta$ with the coupling phase difference $\theta = \pi/2$ and the accumulated phase $\phi_0 = \pi/2$. The inset depicts $R$ versus $\Delta$ with $\theta = \pi/2$ and $\phi_0 = \pi$ in the case of photon propagation time $\tau\Gamma = 1$ and atomic dissipation rate $\gamma_e/\Gamma = 0$, where $\Gamma$ is the atomic emission rate into the waveguide.

The yellow dot-dashed, red dotted, and blue dashed lines in Fig. 3d depict the contrast ratio

$$I = \frac{T_{2\to1} - T_{1\to2}}{T_{2\to1} + T_{1\to2}} \quad (9)$$

versus the coupling phase difference $\theta$ with different atomic dissipation rates. The observed phase-dependent nonreciprocal transmission can be easily understood by resorting to Eqs. (6a) and (7a). For instance, it is viable to have $t = 0$ ($t' = 0$) in the case of $\Delta = 2\Gamma\sin\phi\cos\theta$ and $\gamma_e = 4\Gamma\sin\phi\sin\theta$ ($\gamma_e = -4\Gamma\sin\phi\sin\theta$). Thus, perfectly nonreciprocal transmission denoted by $I = \pm1$ can be attained by tuning the accumulated phase $\phi$ and the coupling phase difference $\theta$ if we have $|\Delta| \leq 2\Gamma$ and $\gamma_e \leq 4\Gamma$. Note, in particular, that the conditions for attaining $I = \pm 1$ will reduce to $\theta = \pi/2 + 2n\pi$ and $\gamma_e = \pm 4\Gamma\sin\phi$ as well as $\theta = -\pi/2 + 2n\pi$ and $\gamma_e = \mp 4\Gamma\sin\phi$ for $\Delta = 0$, with $n$ being an arbitrary integer.

Furthermore, the underlying physics of the reciprocal and nonreciprocal scatterings can be understood via examining the atomic excitation by the single photon. To this end, we define the contrast ratio $D$ of the atomic excitation probabilities for two opposite propagating directions as

$$D = \frac{|u_{e_{2\to1}}|^2 - |u_{e_{1\to2}}|^2}{|u_{e_{2\to1}}|^2 + |u_{e_{1\to2}}|^2} \quad (10)$$

with

$$u_{e_{1\to2}} = \frac{t-1}{-i\frac{g}{v_g}[e^{i\theta_1} + e^{i(\theta_2+\phi)}]},$$
$$u_{e_{2\to1}} = \frac{t'-1}{-i\frac{g}{v_g}[e^{i\theta_2} + e^{i(\theta_1+\phi)}]}. \quad (11)$$

According to Eqs. (6a) and (7a), parameters $t-1$ and $t'-1$ have the same denominator containing $\gamma_e$ but different numerators without $\gamma_e$. Furthermore, because the denominator that contains $\gamma_e$ is eliminated when calculating Eq. (10), the contrast ratio $D$ is independent of dissipation rate $\gamma_e$. Note that the contrast ratio $D$ can be used to capture the difference of the atomic excitation probabilities for opposite directions even if the eigenstate Eq. (2) is unnormalized. It is also not difficult to find from Eq. (6a) that $t = 1$ and hence $u_{e_{1\to2}} = 0$ in the case of $1 + \cos(\phi - \theta) = 0$ while from Eq. (7a) that $t' = 1$ and hence $u_{e_{2\to1}} = 0$ in the case of $1 + \cos(\phi + \theta) = 0$. Then, with $\phi \mp \theta = (2n+1)\pi$, we can attain $D = \pm 1$ as a measure of the optimal difference of atomic excitation probabilities for oppositely propagating photons.

We plot in Fig. 3d the contrast ratio $D$ (black solid line) as a function of the phase difference $\theta$ with $\phi_0 = \pi/2$. For $\theta = \pi/2$, $D = -1$ means that the atom can only be excited by the left-incident photon, and thus the atom-waveguide interaction becomes ideally equivalent chiral[69,71]. In this case, the right-incident photon is guided transparently because it does not interact with the atom. While in the Markovian regime, the reflections are lacking for both directions under the ideally equivalent chiral coupling[18,46], this is in fact not true in the non-Markovian regime as will be discussed in the "Non-Markovian regime" subsection. For $\theta = 3\pi/2$, $D = 1$ corresponds to the ideally equivalent chiral case where the atom can only be excited by the right-incident photon. For other cases of $D = 0$ and $0 < |D| < 1$, the equivalent atom-waveguide couplings are nonchiral and nonideal chiral, respectively. In fact, the nonreciprocal scatterings arise from the different dissipations into the environment, i.e., the energy loss into the environment is proportional to the dissipation rate $\gamma_e$ as well as the atomic population. In addition, we demonstrate in Fig. 3e that the contrast ratio $D$ is also sensitive to the propagating phase. This provides an alternative way to tune the equivalent chirality of the atom-waveguide interaction and the reciprocal/nonreciprocal scattering on demand. Note that the equivalent chiral coupling found here is not in the standard form featuring different coupling strengths[19,43], but is a direct result of asymmetric interference effects between the left- and right-incident photons.

The results above can also be interpreted from the aspect of Hermitian and non-Hermitian scattering centers[72–74]. In our system with $\gamma_e = 0$ ($\gamma_e \neq 0$), the giant atom can be regarded as a Hermitian (non-Hermitian) scattering center of the Aharonov-Bohm structure supporting two spatial interference paths. For the Hermitian case, the scattering remains reciprocal; however, when introducing an imaginary potential, e.g., the external atomic dissipation, the combination of the non-Hermiticity and the broken time-reversal symmetry gives rise to nonreciprocal scatterings[72,73]. It is noted that, as discussed in the case in Fig. 3b, c ($\theta = \pi$), not all non-Hermitian scattering centers can demonstrate nonreciprocal transmissions. The exceptions include, e.g., $\mathcal{P}$-, $\mathcal{T}$-, or $\mathcal{PT}$-symmetric scattering centers[72,74]. In our model, although the giant atom can exhibit chiral spontaneous emission corresponding to the time-reversal symmetry breaking if $\theta \neq n\pi$[69], the scatterings are still reciprocal unless the additional non-Hermiticity (such as external dissipations) are introduced. Similar equivalent asymmetric couplings are also observed in the setup of emitters coupled to photonic lattice[75].

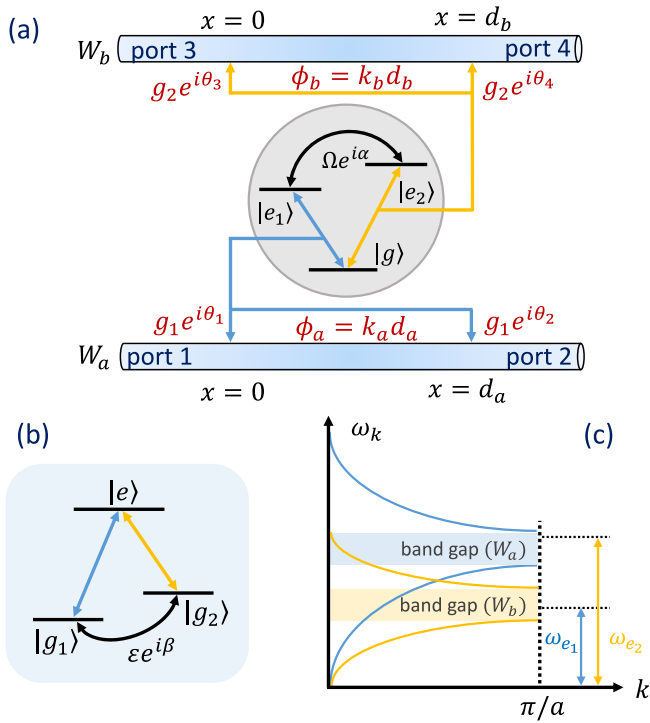

**Fig. 5 Schematic configuration of the three-level giant atom coupled to double waveguides. a** ∇-type atom: the waveguide $W_a$ ($W_b$) is coupled to the transition $|g\rangle \leftrightarrow |e_1\rangle$ ($|g\rangle \leftrightarrow |e_2\rangle$) at two separated points. The excited states $|e_1\rangle$ and $|e_2\rangle$ are coupled to a coherent field $\Omega e^{i\alpha}$. **b** Δ-type atom: $W_a$ ($W_b$) is coupled to $|g_1\rangle \leftrightarrow |e\rangle$ ($|g_2\rangle \leftrightarrow |e\rangle$) at two separated points. Two ground states $|g_1\rangle$ and $|g_2\rangle$ are coupled to a coherent field $\varepsilon e^{i\beta}$. **c** Band structures of two PCWs used to implement our proposals.

The nonreciprocal scatterings can also be observed as the external decay rate is included for a two-level chiral giant atom coupled to a waveguide with asymmetric coupling strengths $g_L \neq g_R$ (see Supplementary Note 1). The problem lies in that it is difficult or impossible to tune the degree of chirality $\eta = g_R/g_L$ in the full (non-periodic) range of $\{0, \infty\}$, e.g., for a special photonic crystal waveguide (PCW) with one side shifted half the lattice constant relative to the other side[44]. In our model, however, it is much easier and more flexible to engineer the nonreciprocal scatterings in a standard PCW by tuning $\phi$ and $\theta$ in the full (periodic) range of $2\{n, (n+1)\}\pi$ via the separation of two coupling points and the magnetic fluxes threading different Josephson junctions[69,76,77], respectively.

*Non-Markovian regime.* With nontrivial local coupling phases, as demonstrated above, the current giant-atom model (in the Markovian regime) is able to simulate a chiral atom-waveguide system. However, one important characteristic of the giant atom is the peculiar scattering behaviors arising in the non-Markovian regime, where the propagating phase accumulation $\phi = \phi_0 + \tau\Delta$ is sensitive to the detuning $\Delta$ due to the large enough $\tau$ that is comparable to or larger than the lifetime of the atom[29]. Such a detuning-dependent phase will undoubtedly result in the non-Markovian features in the transmission and reflection spectra[25,66]. Here we just consider our system in the non-Markovian regime and demonstrate the reflection with $\phi_0 = \pi/2$ and $\theta = \pi/2$. Note that the reflection is totally prevented in the small-atom case with an ideal chiral coupling, which has been demonstrated in ref. [18].

We plot in Fig. 4 the reflection rates $R = |r|^2$ for the left-incident photon in the Markovian and non-Markovian regimes.

The yellow solid curve shows that the reflection in the Markovian regime disappears completely. Such a reflectionless behavior occurs in the case of $D = \pm 1$, independent of the external atomic dissipation. However, in the non-Markovian regime, due to the Δ-dependent propagating phase $\phi$, the reflection revives with multiple peaks aligning periodically in the frequency domain. In addition, the maximums of the reflection peaks decrease gradually with the increasing of $\gamma_e$. The underlying physics is that, in the phase accumulation $\phi$, the non-Markovian contribution $\tau\Delta$ cannot be ignored relative to $\phi_0$; thus, $\tau\Delta$ and $\phi_0$ determine the scattering behaviors jointly. The reflection disappears at some discrete Δ points satisfying $\tau\Delta = n\pi$.

Moreover, we can find from Eq. (6b) the condition

$$(\Delta - 2\Gamma \sin\phi \cos\theta)^2 + \gamma_e^2/4 + 2\gamma_e\Gamma(1 + \cos\theta\cos\phi) + 4\Gamma^2\sin^2\phi\sin^2\theta = 0, \quad (12)$$

for achieving the perfect reflection ($R = 1$). It is easy to see that this equation has no solutions for any choices of $\Delta/\Gamma$, $\theta$, and $\phi$ in the case of $\gamma_e \neq 0$. Hence, it is viable to achieve the perfect reflection only by simultaneously requiring $\sin\phi \sin\theta = 0$ and $\Delta = 2\Gamma \sin\phi \cos\theta$ in the case of $\gamma_e = 0$, which have solutions (i) $\phi = n\pi$ and $\Delta = 0$; (ii) $\theta = 2n\pi$ and $\Delta = 2\Gamma \sin\phi$; (iii) $\theta = (2n+1)\pi$ and $\Delta = -2\Gamma \sin\phi$. The underlying physics is that constructive interference occurs between two reflection paths contributed by coupling points $x = 0$ and $x = d$, respectively, which becomes perfect (imperfect) in the case of $\gamma_e = 0$ ($\gamma_e \neq 0$) due to a vanishing (nonzero) possibility for losing photons via the external decay. In the inset of Fig. 4, we plot $R$ against $\Delta/\Gamma$ for $\theta = \pi/2$ and $\phi_0 = \pi$ as an example and find that $R = 1$ at the resonance point ($\Delta = 0$) as predicted by solution (i).

**Three-level giant atom coupled to double waveguides**. In this section, we consider two types of three-level giant atoms to explore the possibility of realizing nonreciprocal scatterings, as well as relevant single-photon router and circulator applications, without the additional non-Hermiticity (i.e., external atomic dissipation). As shown in Fig. 5a, we propose a ∇-type giant atom coupled to two single-mode waveguides via different transitions sharing the same ground state and driven by an external field on the third transition between two excited states. Figure 5b shows instead a Δ-type giant atom whose two ground states, interacting with an external field, are further coupled to the same excited state via different waveguide modes. The two proposals can be implemented with a three-level transmon coupled to two PCWs by considering that the energy bandgaps of different PCWs don't overlap each other. In this case, it is viable to assume that only one transition driven by the external field exhibits a frequency falling outside the bandgaps of both PCWs, while the frequencies of other two transitions fall outside the bandgaps of different PCWs as shown in Fig. 5c. It is then justified that no atomic transitions will be coupled to both PCWs as long as the incident photons are not resonant with the transition driven by an external field. Experimentally, the upper and lower edges of a PCW's bandgap may be controlled by adding defects[78], adjusting the waveguide widths[79], and varying widths of permalloy and cobalt stripes in crystals[80], while the transition frequencies of a transmon can be tuned via the external flux of a magnetic coil[11].

As shown in Fig. 5a, the atomic transition $|g\rangle \leftrightarrow |e_1\rangle$ of frequency $\omega_{e_1}$ is coupled to waveguide $W_a$ with complex coupling coefficient $g_1 e^{i\theta_{1,2}}$ at two separated points $x = 0$ and $x = d_a$, respectively; the transition $|g\rangle \leftrightarrow |e_2\rangle$ of $\omega_{e_2}$ is coupled to $W_b$ with $g_2 e^{i\theta_{3,4}}$ at $x = 0$ and $x = d_b$, respectively. The excited states $|e_{1,2}\rangle$ are coupled to an external coherent field of Rabi frequency $\Omega$ and initial phase $\alpha$. The atom is initialized in the ground state $|g\rangle$. The

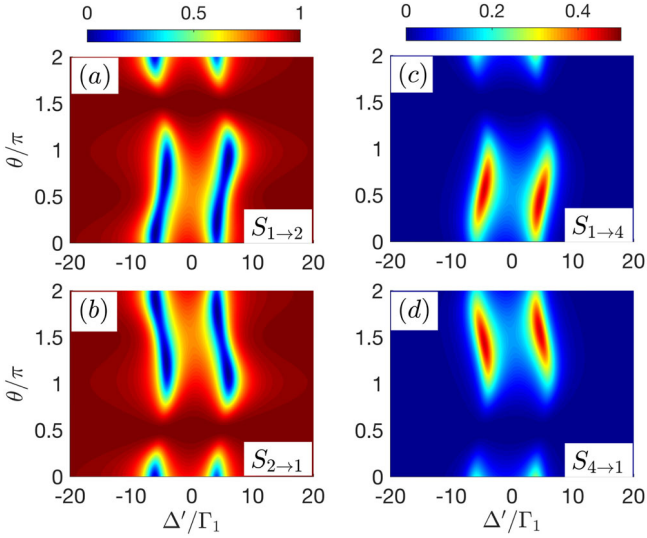

**Fig. 6 Nonreciprocal scattering behaviors in a three-level ∇-type giant-atom system.** Scattering probabilities **a** $S_{1\to 2}$; **b** $S_{2\to 1}$; **c** $S_{1\to 4}$; **d** $S_{4\to 1}$ versus the detuning $\Delta'$ and the coupling phase difference $\theta$. Other parameters: atomic dissipation rates $\gamma_{e_1} = \gamma_{e_2} = 0$, the ratio of two atomic emission rates into different waveguides $\Gamma_2/\Gamma_1 = 1$, Rabi frequency of the external coherent field $\Omega = 5\Gamma_1$, and accumulated phase of the photon traveling in waveguide $W_a$ $\phi_{a0} = \pi/2$.

Hamiltonian of the ∇-type giant atom coupled to two waveguides can be written as

$$H' = H'_w + H'_a + H'_I,$$
$$H'_w = \int_{-\infty}^{+\infty} dx \left[ a_L^\dagger(x)\left(\omega_0 + iv_g \frac{\partial}{\partial x}\right)a_L(x) \right.$$
$$\left. + a_R^\dagger(x)\left(\omega_0 - iv_g \frac{\partial}{\partial x}\right)a_R(x) \right]$$
$$+ \int_{-\infty}^{+\infty} dx \left[ b_L^\dagger(x)\left(\omega_0 + iv_g \frac{\partial}{\partial x}\right)b_L(x) \right.$$
$$\left. + b_R^\dagger(x)\left(\omega_0 - iv_g \frac{\partial}{\partial x}\right)b_R(x) \right],$$
$$H'_a = \left(\omega_{e_1} - i\frac{\gamma_{e_1}}{2}\right)|e_1\rangle\langle e_1| + \left(\omega_{e_2} - i\frac{\gamma_{e_2}}{2}\right)|e_2\rangle\langle e_2|$$
$$+ (\Omega e^{i\alpha}|e_1\rangle\langle e_2| + \text{H.c.}),$$
$$H'_I = \int_{-\infty}^{+\infty} dx \{\delta(x)g_1 e^{i\theta_1} \left[a_R^\dagger(x) + a_L^\dagger(x)\right]|g\rangle\langle e_1|$$
$$+ \delta(x-d)g_1 e^{i\theta_2}\left[a_R^\dagger(x) + a_L^\dagger(x)\right]|g\rangle\langle e_1|$$
$$+ \delta(x)g_2 e^{i\theta_3}\left[b_R^\dagger(x) + b_L^\dagger(x)\right]|g\rangle\langle e_2|$$
$$+ \delta(x-d)g_2 e^{i\theta_4}[b_R^\dagger(x) + b_L^\dagger(x)]|g\rangle\langle e_2| + \text{H.c.}\},$$
(13)

where $a_{R,L}/b_{R,L}$ ($a_{R,L}^\dagger/b_{R,L}^\dagger$) annihilates (creates) right-going and left-going photons in the waveguide $W_a/W_b$, respectively. In the single-excitation subspace, the eigenstate of the system can be expressed as

$$|\Psi\rangle = \int_{-\infty}^{+\infty} dx \left[ \Phi_{aR}(x) a_R^\dagger(x) + \Phi_{aL}(x) a_L^\dagger(x) \right.$$
$$\left. + \Phi_{bR}(x) b_R^\dagger(x) + \Phi_{bL}(x) b_L^\dagger(x) \right] |0, g\rangle$$
$$+ u_{e_1}|0, e_1\rangle + u_{e_2}|0, e_2\rangle,$$
(14)

where $\Phi_{aR,aL}(\Phi_{bR,bL})$ are the probability amplitudes of creating the right-going and left-going photons in $W_a(W_b)$, respectively.

Assuming that a photon with renormalized wave vector $k_a$ is emanated from port 1 of $W_a$, the probability amplitudes can be written as

$$\Phi_{aR}(x) = e^{ik_a x}\{\Theta(-x) + M[\Theta(x) - \Theta(x - d_a)]$$
$$+ s_{1\to 2}\Theta(x - d_a)\},$$
$$\Phi_{aL}(x) = e^{-ik_a x}\{s_{1\to 1}\Theta(-x) + N[\Theta(x) - \Theta(x - d_a)]\},$$
$$\Phi_{bR}(x) = e^{ik_b x}\{Q[\Theta(x) - \Theta(x - d_b)] + s_{1\to 4}\Theta(x - d_b)\},$$
$$\Phi_{bL}(x) = e^{-ik_b x}\{s_{1\to 3}\Theta(-x) + W[\Theta(x) - \Theta(x - d_b)]\},$$
(15)

where the wave vectors $k_a = (E' - \omega_0)/v_g$ with the eigenenergy $E'$ in $W_a$ and $k_b = k_a + (\omega_{e_2} - \omega_{e_1})/v_g$ in $W_b$. When excited to state $|e_1\rangle$ by the incident photon from port 1, the atom can either re-emit a photon with the same frequency to $W_a$ via decaying back to state $|g\rangle$ directly, or radiate a photon with frequency $\omega_{e_2}$ to $W_b$ via first transferring from state $|e_1\rangle$ to state $|e_2\rangle$ due to the external driving and then decaying to state $|g\rangle$[67,81]. If a photon with renormalized wave vector $k_b$ is sent from port 4 of $W_b$, the probability amplitudes can be written as

$$\Phi_{aR}(x) = e^{ik_a x}\{s_{4\to 2}\Theta(x - d_a) + M'[\Theta(x) - \Theta(x - d_a)]\},$$
$$\Phi_{aL}(x) = e^{-ik_a x}\{N'[\Theta(x) - \Theta(x - d_a)] + s_{4\to 1}\Theta(-x)\},$$
$$\Phi_{bR}(x) = e^{ik_b x}\{Q'[\Theta(x) - \Theta(x - d_b)] + s_{4\to 4}\Theta(x - d_b)\},$$
$$\Phi_{bL}(x) = e^{-ik_b x}\{\Theta(x - d_b) + W'[\Theta(x) - \Theta(x - d_b)]$$
$$+ s_{4\to 3}\Theta(-x)\}.$$
(16)

By solving the stationary Schrödinger equation, one can obtain the scattering amplitudes of ∇-type giant atom for this case.

*Nonreciprocal scattering.* For simplicity, we start by supposing $\phi_b = \theta_3 = \theta_4 = 0$, i.e., the transition $|g\rangle \leftrightarrow |e_2\rangle$ is coupled to $W_b$ at a single point. Then, the scattering probabilities can be calculated from $S_{1\to 2} = |s_{1\to 2}|^2$ and $S_{1\to 3(4)} = |s_{1\to 3(4)}|^2$ with the scattering amplitudes given by

$$s_{1\to 2} = \frac{\Delta' + i\frac{\gamma_{e_1}}{2} - \Omega^2 f - 2\Gamma_1 e^{i\theta}\sin\phi_a}{\Delta' + i\frac{\gamma_{e_1}}{2} - \Omega^2 f + 2i\Gamma_1(1 + e^{i\phi_a}\cos\theta)},$$
$$s_{1\to 3(4)} = \frac{g_2 \Omega e^{-i\alpha}(s_{1\to 2} - 1)}{g_1(\Delta' + i\gamma_{e_2}/2 + i\Gamma_2)[e^{i\theta_1} + e^{i(\theta_2 - \phi_a)}]},$$
(17)

with the detuning $\Delta' = E' - \omega_{e_1}$ and the atomic emission rates $\Gamma_{1,2} = g_{1,2}^2/v_g$. As discussed above, we make the substitution $\phi_a = (k_a + k_0)d_a \simeq \phi_{a0}$ in the Markovian regime. Likewise, one can also obtain $S_{2\to 1} = |s_{2\to 1}|^2$ and $S_{2\to 3(4)} = |s_{2\to 3(4)}|^2$ with

$$s_{2\to 1} = \frac{\Delta' + i\frac{\gamma_{e_1}}{2} - \Omega^2 f - 2\Gamma_1 e^{-i\theta}\sin\phi_a}{\Delta' + i\frac{\gamma_{e_1}}{2} - \Omega^2 f + 2i\Gamma_1(1 + e^{i\phi_a}\cos\theta)},$$
$$s_{2\to 3(4)} = \frac{g_2 \Omega e^{-i\alpha}(s_{2\to 1} - 1)}{g_1(\Delta' + i\gamma_{e_2}/2 + i\Gamma_2)[e^{i\theta_2} + e^{i(\theta_1 - \phi_a)}]},$$
(18)

which are achieved via exchanging $\theta_1$ and $\theta_2$ in Eq. (17). It is found that $s_{4\to 1} = s_{2\to 3}$ and thus $S_{4\to 1} = |s_{4\to 1}|^2 = S_{2\to 3}$.

Compared with Eq. (6a) and Eq. (7a) of the two-level giant atom, both $s_{1\to 2}$ and $s_{2\to 1}$ include an additional coupling term $\Omega^2 f$ with

$$f = \frac{1 - i\left(\frac{\gamma_{e_2}}{2} + \Gamma_2\right)}{\Delta'^2 + \left(\frac{\gamma_{e_2}}{2} + \Gamma_2\right)^2},$$
(19)

which describes the photon transfer from $W_a$ to $W_b$. It can be seen from Eqs. (17)–(19) that, in contrast to the two-level giant-atom scheme, the transmission between ports 1 and 2 in $W_a$ is nonreciprocal even if the external dissipations are not considered. In fact, for $\gamma_{e_1} = \gamma_{e_2} = 0$, the imaginary part of $f$ describing the decay of $|e_2\rangle \to |g\rangle$ into $W_b$ plays the role of an external dissipation for the transition $|e_1\rangle \to |g\rangle$.

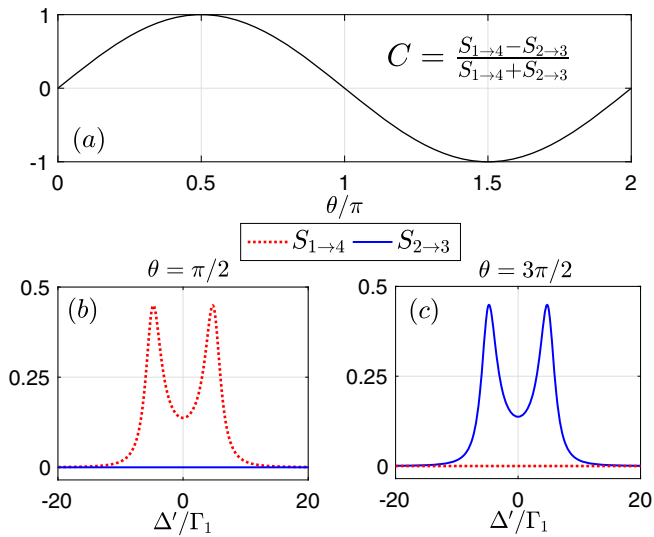

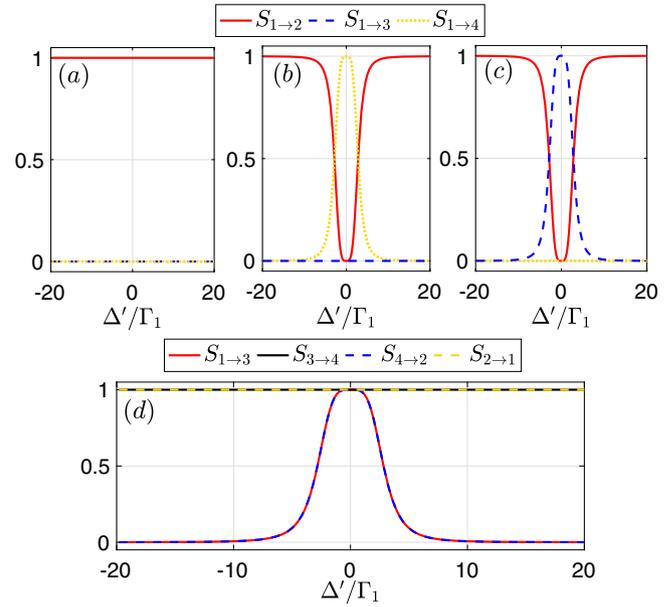

**Fig. 7 Chiral scattering behaviors in a three-level ∇-type giant-atom system. a** Chirality $C$ versus the coupling phase difference $\theta$ with the detuning $\Delta' = 0$. Scattering probabilities $S_{1\to4}$ and $S_{2\to3}$ versus $\Delta'$ for **b** $\theta = \pi/2$ and **c** $\theta = 3\pi/2$. Other parameters: atomic dissipation rates $\gamma_{e_1} = \gamma_{e_2} = 0$, the ratio of two atomic emission rates into different waveguides $\Gamma_2/\Gamma_1 = 1$, Rabi frequency of the external coherent field $\Omega = 5\Gamma_1$, and accumulated phase of the photon traveling in waveguide $W_a$ $\phi_{a0} = \pi/2$.

It is worth noting that the scattering probabilities of the ∇-type system are independent of the phase $\alpha$ of the external coherent field $\Omega$ in spite of the closed-loop atom-level structure. This is because the ∇-type atom cannot provide the inner two-path quantum interference. For instance, when excited to state $|e_1\rangle$ by an incident photon from port 1, the atom may be pumped to state $|e_2\rangle$ by the external field $\Omega$ and then return to state $|g\rangle$ after emitting a photon into $W_b$, which is the only path for the photon transferring from $W_a$ to $W_b$. This is radically different from the Δ-type structure as will be discussed in the "Comparison with the Δ-type scheme" subsection. In fact, the photon cannot be routed from $W_a$ to $W_b$ in the absence of the field $\Omega$, implying that the ∇-type three-level giant atom reduces to a two-level one. This is also consistent with the fact that $S_{1\to3(4)} = S_{2\to3(4)} = 0$ when $\Omega = 0$.

Figure 6 shows the single-photon scattering spectra as functions of the detuning $\Delta'$ and the phase difference $\theta$. As discussed above, it can be seen from Fig. 6a, b that the nonreciprocal scattering can still be realized in $W_a$ ($S_{1\to2} \neq S_{2\to1}$) with $W_b$ playing the role of the external thermal reservoir in the two-level scheme as analyzed above. According to the conclusion in the "Reciprocal and nonreciprocal transmissions" subsection, for $\theta \neq n\pi$ and $\phi_{a0} = \pi/2 + 2n\pi$, the excitation probabilities $|u_{e_1}|^2$ for two opposite directions are unequal, i.e., the interaction between the atom and $W_a$ is equivalent chiral. Then, as shown in Fig. 6c, d, the nonreciprocal scattering between ports 1 and 4 can be led to by the equivalent chiral coupling, since the scattering probability $S_{1\to4}$ ($S_{4\to1}$) is related to the coupling between the atomic transition $|e_1\rangle \leftrightarrow |g\rangle$ and the right-going (left-going) mode in $W_a$. When $\theta = \pi/2$ ($3\pi/2$), $S_{1\to4}$ ($S_{4\to1}$) approaches 0.5 and $S_{4\to1}$ ($S_{1\to4}$) falls to 0. This corresponds to the ideally equivalent chiral case where the atom is only coupled to the right-going (left-going) modes effectively in $W_a$. When $\theta = \pi$, the scatterings between ports 1 and 4 are reciprocal, similar to the results of the equivalent nonchiral case in the "Reciprocal and nonreciprocal transmissions" subsection.

**Fig. 8 Targeted routing and circulating scheme with a three-level ∇-type giant-atom system.** Scattering probabilities versus the detuning $\Delta'$ with different Rabi frequencies of the external coherent field $\Omega$ and coupling phase differences $\theta$ and $\theta'$ **a** $\Omega = 0$, $\theta = \pi/2$; **b** $\Omega = 2\Gamma_1$, $\theta = \pi/2$, $\theta' = \pi/2$; **c** and **d** $\Omega = 2\Gamma_1$, $\theta = \pi/2$, $\theta' = 3\pi/2$. Other parameters: atomic dissipation rates $\gamma_{e_1} = \gamma_{e_2} = 0$, the ratio of two atomic emission rates into different waveguides $\Gamma_2/\Gamma_1 = 1$, and accumulated phases of the photon traveling in waveguides $W_a$ and $W_b$ $\phi_{a0} = \phi_{b0} = \pi/2$.

*Chiral scattering.* Next, we turn to study another kind of asymmetric scattering phenomenon proposed recently called "chiral scattering". Specifically, the transmission from port 1 to port 4 and that from port 2 to port 3 are different. Quantitatively, the chiral scattering can be evaluated by the chirality defined as[82]

$$C = \frac{S_{1\to4} - S_{2\to3}}{S_{1\to4} + S_{2\to3}}. \qquad (20)$$

Figure 7a shows the chirality as a sinusoidal function of the phase difference $\theta$. In view of this, chiral scatterings can be observed as long as $\theta \neq n\pi$, where the chirality $C \neq 0$ means $S_{1\to4} \neq S_{2\to3}$. This can be further verified by the scattering spectra as shown in Fig. 7b, c. Note that $C = 1$ ($C = -1$) corresponding to $\theta = \pi/2$ ($\theta = 3\pi/2$), implies that only the scattering from port 2 (1) to port 3 (4) is prevented, as shown in Fig. 7b [Fig. 7c].

The underlying physics of the chiral scattering can also be attributed to the difference between the atomic excitation probabilities for two incident directions as discussed above. The excitation probabilities $|u_{e_1}|^2$ by the photon incident from port 1 and port 2 can be unequal, and thus the atom is pumped from $|e_1\rangle$ to $|e_2\rangle$ with unequal probabilities. This leads to different probabilities of routing photons from $W_a$ to $W_b$. Furthermore, as shown in Fig. 7, the chiral scattering scheme here shows the in-situ tunability that the scattering chirality can be controlled by tuning the phase difference $\theta$.

*Targeted router and circulator.* In this subsection, we would like to demonstrate how to realize a single-photon targeted router and circulator based on the asymmetric scatterings above. Specifically, one can send a single photon deterministically from port 1 to one of the other three ports on demand. Note that the router and circulator can run with very high efficiency in such a non-loss system. Here we assume the transition $|e_2\rangle \leftrightarrow |g\rangle$ coupled to $W_b$ at two separated points, i.e., $\phi_b \neq 0$, as shown in Fig. 5a, and define $\theta' = \theta_4 - \theta_3$.

The mechanism of the targeted router can be understood from Fig. 8a–c showing the scattering probabilities from port 1 to other ports versus the detuning $\Delta'$. When turning off the external field ($\Omega \equiv 0$), as shown in Fig. 8a, the incident photon from port 1 cannot be routed to $W_b$; particularly for $\theta = \pi/2$, the photon is routed to port 2 totally. Next, we turn on the external field to enable photon routing to the desired port in $W_b$ with high efficiency. When setting $\theta' = \pi/2$, as shown in Fig. 8b, a photon resonant with the transition $|g\rangle \leftrightarrow |e_1\rangle$ can be routed from port 1 to port 4 totally. Likewise, when setting $\theta' = 3\pi/2$ as shown in Fig. 8c, the resonant photon can be routed to port 3 totally. In addition, both the propagating phases $\phi_{a0}$ and $\phi_{b0}$ determine the output port of photons in $W_b$, which is a unique feature of the giant-atom model.

It is worth noting that one $\nabla$-type small atom with chiral asymmetric couplings ($g_{1L} \neq g_{1R}$ and $g_{2L} \neq g_{2R}$) to two waveguides has also been explored to realize a deterministic routing[19], with subscripts "1,2" referring to the first and second waveguides while "L,R" to the left- and right-moving photons, respectively. In principle, it is viable to observe any desired routing results by tailoring two degrees of chirality $\eta_1 = g_{1R}/g_{1L}$ and $\eta_2 = g_{2R}/g_{2L}$, e.g., via the amplitude of a magnetic field applied upon a quantum dot serving as the small atom[44]. The problem is that $\eta_1$ and $\eta_2$ exhibit similar changing trends and are located at a single point, hence cannot be tuned independently. In our giant-atom model, however, it is much easier to observe different routing results by tailoring $\theta$ and $\theta'$ individually, e.g., via the magnetic fluxes threading Josephson junctions at different coupling points. Such a selective tunability of coupling phase differences can also be used to realize a perfect circulator as shown below, which is impracticable yet by tailoring $\eta_1$ and $\eta_2$. Our giant-atom model bears also another nontrivial feature - the non-Markovian retardation effect, which could result in multiple peaks in the reflection (and also transmission) spectra as shown in Fig. 4, and might enable the simultaneous manipulation of more than one incident photon with different frequencies.

More interestingly, the $\nabla$-type giant atom is also a promising candidate of realizing a single-photon circulator. When turning on the external field and setting $\theta = \pi/2$ and $\theta' = 3\pi/2$, the two waveguides are coupled to the atom with ideally equivalent chiral couplings in opposite manners, respectively. That is to say, the atom is only coupled to the left-incident photons in $W_a$ yet to right-incident photons in $W_b$. Then, as shown in Fig. 8d, one has $S_{2 \to 1} = S_{3 \to 4} \equiv 1$ over the whole frequency range and $S_{1 \to 3} = S_{4 \to 2} = 1$ around the resonance. Consequently, for a resonant photon, directional scattering along the direction $1 \to 3 \to 4 \to 2 \to 1$ can be realized suggesting a high-performance single-photon circulation scheme for quantum networks[59,60].

*Comparison with the $\Delta$-type scheme.* Finally, we consider a $\Delta$-type giant-atom scheme where the $\nabla$-type atom in Fig. 5a is replaced by a $\Delta$-type one in Fig. 5b and compare the single-photon scatterings of these two schemes. The $\Delta$-type structure is constructed with an external coherent filed $\varepsilon e^{i\beta}$ which couples the two ground states $|g_{1,2}\rangle$ of a $\Lambda$-type atom that has been broadly studied to demonstrate quantum interference phenomena, such as coherent population trapping[83] and electromagnetically induced transparency[84].

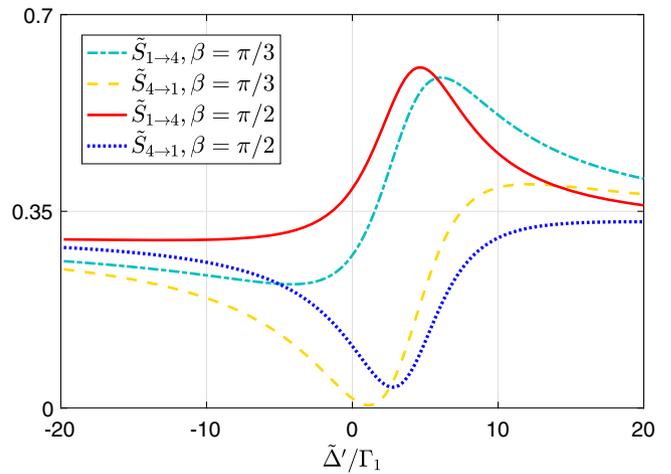

**Fig. 9 Nonreciprocal scattering behaviors in a three-level $\Delta$-type giant-atom system.** Scattering probabilities versus the detuning $\tilde{\Delta}'$ with different phases of the external coherent filed $\beta$. Other parameters: atomic dissipation rates $\gamma_{e_1} = \gamma_{e_2} = 0$, the ratio of two atomic emission rates into different waveguides $\Gamma_2/\Gamma_1 = 1$, Rabi frequency of the external coherent field $\varepsilon = 30\Gamma_1$, and accumulated phases of the photon traveling in waveguides $W_a$ and $W_b$ $\phi_{a0} = \phi_{b0} = \pi/2$.

For the $\Delta$-type giant-atom system, the Hamiltonians of the atom and the atom-waveguide interaction become

$$\begin{aligned} H'_a &= \left(\omega_{g_2} - i\frac{\gamma_{g_2}}{2}\right)|g_2\rangle\langle g_2| + \left(\omega_e - i\frac{\gamma_e}{2}\right)|e\rangle\langle e| \\ &\quad + (\varepsilon e^{i\beta}|g_1\rangle\langle g_2| + \text{H.c.}), \\ H'_I &= \int_{-\infty}^{+\infty} dx \{\delta(x)g_1 e^{i\theta_1}\left[a_R^\dagger(x) + a_L^\dagger(x)\right]|g_1\rangle\langle e| \\ &\quad + \delta(x-d)g_1 e^{i\theta_2}\left[a_R^\dagger(x) + a_L^\dagger(x)\right]|g_1\rangle\langle e| \\ &\quad + \delta(x)g_2 e^{i\theta_3}\left[b_R^\dagger(x) + b_L^\dagger(x)\right]|g_2\rangle\langle e| \\ &\quad + \delta(x-d)g_2 e^{i\theta_4}\left[b_R^\dagger(x) + b_L^\dagger(x)\right]|g_2\rangle\langle e| + \text{H.c.}\} \end{aligned} \quad (21)$$

The single-excitation eigenstate of the system takes the form

$$\begin{aligned} |\Psi\rangle &= \int_{-\infty}^{+\infty} dx \Big\{ \left[\Phi_{aR}(x)a_R^\dagger(x) + \Phi_{aL}(x)a_L^\dagger(x)\right]|0,g_1\rangle \\ &\quad + \left[\Phi_{bR}(x)b_R^\dagger(x) + \Phi_{bL}(x)b_L^\dagger(x)\right]|0,g_2\rangle \Big\} + u_e|0,e\rangle. \end{aligned} \quad (22)$$

With the same procedure above (see Supplementary Note 2), one can obtain the scattering probabilities in this case.

Setting the atom in the ground state $|g_1\rangle$ initially, we plot in Fig. 9 the scattering spectra of $\tilde{S}_{1 \to 4}$ and $\tilde{S}_{4 \to 1}$. It is worth noting that, even in the absence of the local coupling phases, i.e., $\theta = \theta' = 0$, the nonreciprocal scatterings still exist. This is obviously distinct from the $\nabla$-type case. The nonreciprocity of the $\nabla$-type case stems from the equivalent chiral couplings owing to the nontrivial coupling phase difference, and is independent of the phase of the external field. For the $\Delta$-type scheme, however, the nonreciprocity arises from the typical which-way quantum interference, i.e., the interference between the two transition paths $|g_1\rangle \to |g_2\rangle$ and $|g_1\rangle \to |e\rangle \to |g_2\rangle$. In this case, the optical responses are typically sensitive to the phase of the external field encoded in the closed-loop level structure[85]. However, the main drawback to the $\Delta$-type scheme is that one cannot switch on/off the photon transfer between the two waveguides by tuning the external field solely.

Our ∇-type and Δ-type giant atoms will reduce to the corresponding small atoms if we set $\phi_a = \phi_b = 0$ (i.e., $d_a = d_b = 0$). In this case, chiral scattering disappears for both ∇-type and Δ-type small atoms due to the intrinsic symmetry of atom-waveguide interactions. On the other hand, nonreciprocal scattering still can be observed for the Δ-type small atom due to the asymmetric interference (for left- and right-incident photons) between two transitions sharing the same starting and ending states, but will not occur for the ∇-type small atom in the presence of only one accompanied transition and thus absence of any interference effects (see Supplementary Note 3).

## Conclusions

In summary, we have investigated step-by-step the conditions of single-photon nonreciprocal and chiral scatterings in the two-level and three-level giant-atom structures with tunable local phase on each atom-waveguide coupling. We found that the atomic excitation in the two-level giant-atom structure depends on the propagation direction of waveguide modes and can be tuned by the nontrivial coupling phase difference. In such scenario, our two-level giant atom in the Markovian regime is equivalent to a two-level small atom chirally coupled to the waveguide mode. However, it is worth noting that the realization of nonreciprocal scatterings requires the combination of the time-reversal symmetry breaking induced by the local coupling phases and the non-Hermiticity induced by the external atomic dissipation due to the surrounding non-waveguide modes. Moreover, in the non-Markovian regime, the reflection spectra exhibit peculiar non-Markovian features with multiple reflection peaks that are absent in the chiral small-atom case.

For exploring more interesting asymmetric scattering properties and applications with such giant-atom structures, we have extended the two-level structure to the three-level ∇-type and Δ-type ones coupled to two waveguides via different atomic transitions. We found that, for the atomic transition coupled to one waveguide, the transition coupled to the other waveguide can serve as the external dissipation channel. Such three-level giant-atom structures coupled to double waveguides enable the nonreciprocal and chiral scatterings without external dissipations. Based on this mechanism, the high-efficiency single-photon targeted router and circulator can be implemented. Finally, we explained the different physical mechanisms that lead to the nonreciprocal and chiral scatterings for the two phase-sensitive closed-loop three-level giant-atom structures. We believe that our results have promising applications in designing effective and efficient single-photon optical elements for quantum network engineering and optical communications.

## Methods

In this theoretical work, the methods used are solving the stationary Schrödinger equation with the Hamiltonian and single-excitation eigenstate (as described in the main text [Eq. (3)] and Supplementary Material).

## Data availability

All data are available in the main text or in the supplementary materials.

## Code availability

The code used to produce the figures in this article is available from the corresponding author upon request.

## Acknowledgements

This work is supported by the National Natural Science Foundation of China (Nos. 12074030, 12074061, U1930402), the National Key Research and Development Program of China (No. 2021YFE0193500), the Fundamental Research Funds for the Central Universities (No. 2412019FZ045), and the Science Foundation of the Education Department of Jilin Province during the 14th Five-Year Plan Period (No. JJKH20211279KJ).

## Author contributions
Y.T.C. conceived the study, performed the calculations, and wrote the first version of the manuscript. L.D., L.G., and Y.Z. checked the calculations, analyzed the data, and redrafted the manuscript. Z.W., Y.L., and J.H.W. helped in the interpretation of the results and in the writing of the manuscript. All authors discussed the results and reviewed the manuscript.

## Competing interests
The authors declare no competing interests.

## Additional information
**Supplementary information** The online version contains supplementary material available at https://doi.org/10.1038/s42005-022-00991-3.

**Correspondence** and requests for materials should be addressed to Yan Zhang, Yong Li or Jin-Hui Wu.

**Peer review information** *Communications Physics* thanks Anton Kockum and the other, anonymous, reviewer(s) for their contribution to the peer review of this work. Peer reviewer reports are available.

**Reprints and permission information** is available at http://www.nature.com/reprints

**Publisher's note** Springer Nature remains neutral with regard to jurisdictional claims in published maps and institutional affiliations.